\documentclass[10pt, a4paper]{article}
\usepackage[english]{babel}
\usepackage{graphicx, pdflscape, float}
\usepackage{amssymb}
\usepackage{xcolor}
\usepackage{multicol}
\usepackage{setspace}
\usepackage[margin=0.75in]{geometry}
\usepackage{amsmath}
\usepackage{amsthm}
\newtheorem*{remark}{Remark}

\providecommand{\keywords}[1]
{\textbf{\textit{Keywords:}}#1}
\begin{document}
\doublespacing
\title{Estimation of the Parameters of Vector Autoregressive (VAR) Time Series Model with Symmetric Stable Noise}
\author{Aastha M. Sathe and N. S. Upadhye\\
Email address: aasthamadonnasathe@gmail.com, neelesh@iitm.ac.in\\
Centre of Excellence in Computational Mathematics and Data Science,\\
Department of Mathematics, Indian Institute of Technology Madras,
Chennai-600036, INDIA}

\date{\today}
\maketitle
\begin{abstract}
\noindent
In this article, we  propose the fractional lower order covariance method (FLOC) for estimating the parameters of vector autoregressive process (VAR) of order $p$, $p\geq 1$ with symmetric stable noise. Further, we show the efficiency, accuracy and simplicity of our methods through  Monte-Carlo simulation.

\noindent
\keywords{ VAR model, stable distributions, parameter estimation, simulation}
\end{abstract}
\section{Introduction}
\label{sec1}
The univariate stationary time series models, namely, the autoregressive models (AR), moving average models (MA) and the general autoregressive moving average models (ARMA) are popular tools in the statistical analysis of univariate time series data (see \cite{hong}-\cite{T}). On the other hand, for the analysis of multivariate time series data, one of the most successful, flexible, and easy to use models is the vector autoregressive (VAR) model. The VAR model is especially useful for describing. In the classical definition, the above mentioned models are assumed to be second-order due to finite second moment of the noise term. However, these models fail to capture the heavy-tails of the data. This motivates us to use the family of stable distributions to model the data. Some of the significant and attractive features of stable distributions, apart from stability are heavy-tails, leptokurtic shape,  domains of attraction, infinite second moment (with the exception of Gaussian case) and skewness. For more details on stable distributions, see \cite{sam}. Hence, there is a need to explore the behaviour of the above mentioned models with stable noise for effective modelling of the time series data which also involves estimation of the parameters of these models. However, because of the infinite variance, only a handful of estimation techniques are available for models based on stable noise (see \cite{mik}-\cite{hill}). 

\noindent
The structure of dependence for the stable-based models cannot be described by the covariance or correlation functions (univariate case) and cross-covariance or cross-correlation (multivariate case). However, one can find alternative measures of dependence that can replace the classical ones in the case of infinite variance. Some of them are: codifference, covariation and fractional lower order covariance (FLOC) for one-dimensional models and cross-codifference, cross-covariation and cross-FLOC  for multidimensional models (see \cite{sundar}-\cite{ros2}).
 
\noindent
In this article, we develop a method for estimating the parameters of multidimensional VAR model of order $p$, $p\geq 1$, with symmetric stable noise for $\alpha \in [1.5, 2]$. The proposed method employs the use of FLOC, considered as an extension of the covariance function to the stable case and with several interesting applications. The method is reasonable and effective both from the theoretical and practical aspects. The efficiency of the method is shown on the simulated data and by comparing it with the classical least squares and Yule-Walker method for VAR models.

\noindent
The paper is organized as follows. Section \ref{sec2} gives a brief introduction to the stable distributions and bidimensional VAR($p$) model along with the necessary definitions and notations. Section \ref{sec3} discusses the FLOC based parameter estimation method. Section \ref{sec4} deals with simulations and comparative analysis of the proposed method with the classical Yule-Walker method. Section \ref{app} deals with an application to financial data. Finally, Section \ref{con} gives some concluding remarks on the proposed method.

\section{Preliminaries and Notations}\label{sec2}
\subsection{Stable Distributions}
These distributions form a rich class of heavy-tailed distributions, introduced by Paul L\'evy \cite{lev}, in his study on the Generalized Central Limit Theorem. In the one-dimensional case, each distribution, in this class, is characterized by four parameters, namely $\alpha$,  $\beta$,  $\sigma$, and $\delta$, which, respectively, denote the index of stability, skewness, scale and shift of the distribution. Their respective ranges are given by  $\alpha \in (0,2]$, $\beta \in [-1, 1]$, $\sigma > 0$ and $\delta \in \mathbb{R}$. For more details, see 
\cite{hybrid}. The  characteristic function representation for a univariate stable random variable $Z$ is given by \cite{sam}
\begin{eqnarray}
\phi( \textit{t})&=& \begin{cases}
\exp\left\{-(\sigma|t|)^\alpha\left[ 1 - \iota \beta{\rm sign}(t)\tan\left(\frac{\pi\alpha}{2}\right)\right]+\iota \delta t\right\}, &~~~~~~~~~~~~~~~~~~~\alpha \neq 1,\\
\exp\left\{-\sigma|t|\left[ 1+ \iota \beta \frac{2}{\pi}{\rm sign}(t)\ln|t|\right]+ \iota \delta \textit{t} \right\},  &~~~~~~~~~~~~~~~~~~~\alpha=1,
\end{cases}\label{1par}
\end{eqnarray}

\noindent
In this paper, we deal with symmetric stable distributions. We say that the distribution is symmetric  around zero if and only if $\beta=\delta=0$ in (\ref{1par}), i.e., if the characteristic function is 
\begin{equation}\label{2par}
\phi(t)=\mathbb{E}(\exp it Z)=\exp\left\{-(\sigma|t|)^\alpha\right\}
\end{equation}
Note that when $\alpha=2$, $\beta=0$ and $\delta=0$, the distribution is $S\alpha S$ Gaussian. Also, except the Gaussian case with $\alpha=2$, the variance of $Z$ is infinite.

\noindent
In the multi-dimensional case, the characteristic function of a symmetric stable vector $\mathbf{Z}=(Z_1,Z_2,\cdots,Z_r)$ takes the following form \cite{sam}
\begin{equation}
\mathbb{E}(\exp i<\boldsymbol{t},\boldsymbol{Z}>)=\exp\Big \lbrace\int_{S_{r }}\mid <\boldsymbol{t},\boldsymbol{s}> \mid^\alpha \Gamma (ds) \Big\rbrace \label{88}
\end{equation}
where $\Gamma(\cdot)$ is a finite spectral symmetric measure on the unit sphere $S_r$ in $\mathbb{R}^r$ and $<\cdot,\cdot>$ is the inner product. Thus, a necessary and sufficient condition for a stable vector to be symmetric is that the shift vector $\delta_{\boldsymbol{0}}=\boldsymbol{0}$ and $\Gamma(\cdot)$ is a finite spectral symmetric measure on $S_r$. Note that the information about skewness and scale of the multi-dimensional stable distributions are included in the spectral measure $\Gamma(\cdot)$. 

\subsection{Multidimensional VAR($p$) model with symmetric stable noise}
Next, we discuss some important definitions and notations required for parameter estimation of  multidimensional vector autoregressive of order $p$ (VAR($p$)) model with symmetric stable noise. We begin our discussion with the classical definitions of the second-order white noise and of the general multidimensional VAR($p$) model which is later extended and modified to incorporate the infinite-variance noise instead of the classical finite variance white noise.

\noindent
Let the multidimensional time series $\lbrace \mathbf{Z}_{t}=(Z_{1t},\cdots, Z_{rt})^T:t\in \mathbb{Z}\rbrace$ be a \textbf{white noise process} with mean $\boldsymbol{0}$ and covariance matrix $\sum$ if $\lbrace \mathbf{Z}_{t}\rbrace$ is weak-sense stationary with mean vector $\boldsymbol{0}$ and covariance matrix function given by \cite{brock}
\begin{eqnarray}
\gamma(h)=\mathbb{E}[\mathbf{Z}_{(t+h)}^T\mathbf{Z}_t]=\begin{cases}
\sum ~~\rm{when}~\textit{h}=0,\\
0 ~~~~\rm{otherwise}.
\end{cases} \label{0par}\\
\end{eqnarray}

\noindent
Let the multidimensional time series $\lbrace \mathbf{X}_{t}=(X_{1t},\cdots, X_{rt})^T:t\in \mathbb{Z}\rbrace$ (mean-corrected) be a causal \textbf{VAR($p$) process} if it is weak-sense stationary and for all $t \in \mathbb{Z}$ it satisfies the following equation (\cite{brock}, \cite{sundar})
\begin{equation}
 \mathbf{X}_{t}=A_1\mathbf{X}_{t-1}+\cdots+A_p\mathbf{X}_{t-p}+\mathbf{Z}_{t}
\end{equation}
where $\lbrace \mathbf{Z}_{t}\rbrace$ is a multidimensional white noise and $A_1\cdots A_p$ are $r\times r$ matrices of the coefficients. Moreover,
\begin{equation}
\det(I-A_1z-\cdots-A_pz^p)\neq 0
\end{equation}
for all $z\in \mathbb{C}$ such that $|z|\leq 1$, where $I$ denotes an identity matrix.\\
Equivalently, if there exists matrices $\lbrace \Psi_j\rbrace$ with absolutely summable components such that for all $t \in \mathbb{Z}$
\begin{equation}
\mathbf{X}_t=\sum_{j=0}^\infty \Psi_j\mathbf{Z}_{t-j}
\end{equation}
where the matrices $\lbrace \Psi_j\rbrace$ are found recursively
\begin{equation}
\Psi_j=\Phi_j+\sum_{k=1}^\infty A_k\Psi_{j-k},~\mathrm{for}~ j=0,1,2,\cdots
\end{equation}
where $\Phi_0=I$, $\Phi_j=0$ for $j>0$, $A_j=0$ for $j>p$, $\Psi_j=0$ for $j<0$.

\noindent
Let the multidimensional time series $\lbrace \mathbf{X}_{t}=(X_{1t},\cdots, X_{rt})^T:t\in \mathbb{Z}\rbrace$ (mean-corrected) be a causal \textbf{VAR($p$) process with symmetric stable noise} if for all $t \in \mathbb{Z}$ it satisfies the following equation \cite{sundar}
\begin{equation}
\mathbf{X}_{t}=A_1\mathbf{X}_{t-1}+\cdots+A_p\mathbf{X}_{t-p}+\mathbf{Z}_{t}\label{89}
\end{equation}
where the multidimensional noise $\lbrace \mathbf{Z}_{t}\rbrace$ is a symmetric stable vector in $\mathbb{R}^r$ with the characteristic function defined in (\ref{88}) and $A_1,\cdots, A_p$ are $r\times r$ matrices of the coefficients. Additionally, we assume that $\mathbf{Z}_{t}$ is independent from $\mathbf{Z}_{t+h}$ for all $h \neq 0$. Note that the causality of the model is defined in the same way as in the classical \textbf{VAR($p$) process} defined above.

\subsection{Measures of dependence for stable processes}
Note that, due to undefined covariance when $\alpha < 2$, the classical dependence mesaures such as the autocovariance or autocorrelation function cannot be considered as a tool for developing methods of parameter estimation of the process defined in (\ref{89}). In such case, alternative measures of dependence are available in literature that can replace the classical dependence measures. A few known choices are: normalized autocovariation, autocodifference and fractional lower order covariance (FLOC). For details, see (\cite{sundar}, Subsection 2.1 ). Amongst these choices, for the estimation of the parameters of the process defined in (\ref{89}), we prefer to use FLOC due to its simple formulation and accuracy. Next, we present the definitions of FLOC and the cross-FLOC estimator.
\subsubsection{Fractional lower order covariance-FLOC}
The fractional lower order covariance for a 
bidimensional symmetric stable random vector $(X,Y)$ is defined as follows \cite{ma}
\begin{equation}
\mathrm{FLOC}(X,Y,A,B)=\mathbb{E}[X^{<A>}Y^{<B>}]\label{90}
\end{equation}
such that $A,B\geq 0$ satisfying $A+B<\alpha$. 

\noindent
Note that the term defined in (\ref{90}) is dependent on the choice of $A$ and $B$ and in the Gaussian case, it reduces to the classical covariance when $A=B=1$. The above measure is applicable to any symmetric stable vector, even with $0<\alpha<1$.

\noindent
Also, we observe that when $A=1,~B=q-1$, where $q \in[1,\alpha)$ and $\alpha \in (1,2)$, the following relation holds between FLOC and covariation function $\mathrm{CV}(X,Y)$ \cite{sundar}
\begin{equation}
\mathrm{FLOC}(X,Y,1,q-1)=\frac{\mathrm{CV}(X,Y)\mathbb{E}(|Y|^q)}{q\sigma_Y^{\alpha}},
\end{equation}
where
 $$\mathrm{CV}(X,Y)=q\frac{\mathbb{E}(XY^{<q-1>})}{\mathbb{E}(|Y|^q)}\sigma_Y^{\alpha},q\in[1,\alpha).$$
\begin{remark}
	Both the measures of dependence for stable processes, namely, $\mathrm{CV}(X,Y)$ and $\mathrm{FLOC}(X,Y,A,B)$ are not symmetric in its arguments as opposed to the classical covariance function $\mathrm{Cov}(X,Y)$.
	\end{remark}
\noindent
The FLOC is used as a measure of the spatio-temporal dependence of the multidimensional process $\lbrace \mathbf{X}_{t}=(X_{1t},\cdots, X_{rt})^T:t\in \mathbb{Z}\rbrace$. The estimator of the cross-FLOC is defined as follows \cite{sundar}
\begin{equation}
\widehat{FLOC}(X_{it},X_{j(t-k)},A,B)=\frac{\sum_{n=L_1}^{L_{2}}|x_{in}|^A|x_{j(n-k)}|^B\mathrm{sign}(x_{in}x_{j(n-k)})}{L_2-L_1},~i,j=1,\cdots,r\label{7}
\end{equation}
where $\lbrace x_{i1},x_{i2},\cdots,x_{iN}\rbrace$ and $\lbrace x_{j1},x_{j2},\cdots,x_{jN}\rbrace$ are sample trajectories of length $N$ corresponding to the multivariate process $( X_{1t},\cdots,X_{rt})$ and $L_1=\max(0,k)$, $L_2=\min(N,N+k)$. Note that in (\ref{7}) replace $k$ with $-k$ to obtain $\widehat{FLOC}(X_{it},X_{j(t+k)},A,B)$

\section{FLOC based parameter estimation of multidimensional VAR($p$) process with symmetric stable noise }\label{sec3}
In this section, we propose a new method for estimating the coefficients of the $r\times r$ matrices $A_1 \cdots A_p$ of a causal multidimensional VAR($p$) process $\lbrace \mathbf{X}_{t}=(X_{1t},\cdots,X_{rt})^T:t\in \mathbb{Z}\rbrace$ with symmetric stable noise. The proposed method is based on FLOC introduced in the previous section. FLOC is well defined for stable distributions and can be used as a substitute of covariance function specially when the second moment is infinite. The proposed method is discussed below.

\begin{itemize}
\item Let  $\lbrace X_{t}\rbrace$ be a multidimensional causal VAR($p$) process with symmetric stable noise given by
\begin{equation}
\mathbf{X}_{t}=A_{1}\mathbf{X}_{t-1}+\cdots+A_{p}\mathbf{X}_{t-p}+\mathbf{Z}_{t} \label{1}
\end{equation}
where $\lbrace \mathbf{Z}_{t} \rbrace$ is a multidimensional symmetric stable noise with $\alpha >1$.
\item Multiply both sides of (\ref{1}) by $ \lbrace\mathbf{X}_{t-l}^{<B>^T}=( X_{1(t-l)}^{<B>},\cdots,X_{r(t-l)}^{<B>})\rbrace$ and taking expectation, we obtain the following expression
\begin{equation}
\mathbb{E}[\mathbf{X}_{t}\mathbf{X}_{t-l}^{<B>^T}]=
A_1\mathbb{E}[\mathbf{X}_{t-1}\mathbf{X}_{t-l}^{<B>^T}]+\cdots+A_p\mathbb{E}[\mathbf{X}_{t-p}\mathbf{X}_{t-l}^{<B>^T}]\label{2}
\end{equation}
\item
Next, define\\
\begin{equation} \mathbf{\Gamma}_l=FLOC(\mathbf{X}_{t},\mathbf{X}_{t-l},1,q-1)=
\mathbb{E}[\mathbf{X}_{t}\mathbf{X}_{t-l}^{<q-1>^T}],~ l=0,\pm 1, \pm 2,\cdots
\end{equation}
where $1\leq q<\alpha$ and $1<\alpha<2$. Thus, $\mathbf{\Gamma}_l$ represents $r\times r$ lag $l$ cross-FLOC matrix given as
$$ 
\mathbf{\Gamma}_l=
\begin{bmatrix}
	\mathbb{E}[X_{1t}X_{1(t-l)}^{<q-1>}]& \dots & \mathbb{E}[X_{1t}X_{r(t-l)}^{<q-1>}] \\
	\vdots &\cdots&\vdots \\
	\mathbb{E}[X_{rt}X_{1(t-l)}^{<q-1>}]& \dots & \mathbb{E}[X_{rt}X_{r(t-l)}^{<q-1>}]
\end{bmatrix}
$$
\item Thus, (\ref{2}) takes the following form
\begin{equation}
\mathbf{\Gamma}_l=A_1\mathbf{\Gamma}_{l-1}+\cdots+A_p\mathbf{\Gamma}_{l-p},~l=1,\cdots p 
\end{equation} 
and we get the following system of matrix equations
$$ 
[\mathbf{\Gamma}_1\cdots\mathbf{\Gamma}_p]=[A_1\cdots A_p]
\begin{bmatrix}
\mathbf{\Gamma}_0 & \mathbf{\Gamma}_1 & \dots & \mathbf{\Gamma}_{p-1} \\
\mathbf{\Gamma}_{-1} & \mathbf{\Gamma}_0 & \dots & \mathbf{\Gamma}_{p-2} \\
\vdots & \vdots &\cdots&\vdots \\
\mathbf{\Gamma}_{1-p} & \mathbf{\Gamma}_{2-p} &\cdots & \mathbf{\Gamma}_0
\end{bmatrix}
$$
\item
Finally, the estimates of the coefficients of the $r\times r$ matrices $A_1 \cdots A_p$ are obtained from the expression given below using the estimator of the cross-FLOC defined in (\ref{7}).
$$[\widehat{A}_1 \cdots\widehat{A}_p]=[\mathbf{\widehat{\Gamma}}_1\cdots\mathbf{\widehat{\Gamma}}_p]
\begin{bmatrix}
\mathbf{\widehat{\Gamma}}_0 & \mathbf{\widehat{\Gamma}}_1 & \dots & \mathbf{\widehat{\Gamma}}_{p-1} \\
\mathbf{\widehat{\Gamma}}_{-1} & \mathbf{\widehat{\Gamma}}_0 & \dots & \mathbf{\widehat{\Gamma}}_{p-2} \\
\vdots & \vdots &\cdots&\vdots \\
\mathbf{\widehat{\Gamma}}_{1-p} & \mathbf{\widehat{\Gamma}}_{2-p} &\cdots & \mathbf{\widehat{\Gamma}}_0
\end{bmatrix}^{-1}
$$ 

\end{itemize}
\section{Simulation and Comparative Analysis}
\label{sec4}
In this section, we investigate the performance of the proposed FLOC based parameter estimation method for bidimensional VAR($p$) models with symmetric stable noise through Monte Carlo simulations. Further, focussing on the practical aspect, we compare the results of the proposed method with the classical least squares (LS) and Yule-Walker (Y-W) \cite{lue} method to emphasize the difference in the two approaches. Note that, $A=1$ is fixed throughout.

\noindent
To test the proposed estimation procedure, we consider the bidimensional VAR(2) model $\lbrace \mathbf{X}_{t}=(X_{1t}, X_{2t})^T:t\in\mathbb{Z}\rbrace $ with bidimensional independent symmetric stable noise $\mathbf{Z}_{t}=(Z_{1t}, Z_{2t})^T:t\in\mathbb{Z}\rbrace $ given below
\begin{equation}
 \mathbf{X}_{t}=A_1\mathbf{X}_{t-1}+A_2\mathbf{X}_{t-2}+\mathbf{Z}_{t}\label{a}
\end{equation}
where $A_1=\begin{bmatrix}
	a_1 & a_3\\
	a_2 & a_4
\end{bmatrix}$ and $A_2=\begin{bmatrix}
a_5 & a_7\\
a_6 & a_8
\end{bmatrix}$.\\
Next, for different values of the sample size $n$, $\alpha \in [1.5,2]$, $A_1=\begin{bmatrix}
0.1 & 0.3\\
0.2 & 0.1
\end{bmatrix}$ and $A_2=\begin{bmatrix}
0.2 & 0.2\\
0.05 & 0.1
\end{bmatrix}$, simulations are run where 500 realisations of the model in (\ref{a}) are generated using the function \textit{varima.sim} available in ``portes" package in R. For each realisation, we calculate $\widehat{A}_1$ and $\widehat{A}_2$ for different values of $B$. From Tables \ref{12}, \ref{11}, \ref{13} and \ref{15}, we observe that for all considered $n$, $\alpha$ and different values of $B$, the root mean squared errors (RMSE) get smaller as $B$ approaches close to $\alpha -1$.

\subsection*{Comparison of classical LS and Y-W with FLOC for VAR models}
For comparison of our proposed method with the classical LS and Y-W method, we use the \textit{VAR.est} function available in the ``VAR.etp" package and \textit{marfit} function available in the ``TSSS" package in R respectively. Next, simulations are run where 500 realisations of the model in (\ref{a}), for different values of $n$ and $\alpha$, are generated using the function \textit{varima.sim} available in ``portes" package in R, where   $A_1=\begin{bmatrix}
0.1 & 0.3\\
0.2 & 0.1
\end{bmatrix}$ and $A_2=\begin{bmatrix}
0.3 & 0.2\\
0.4 & 0.1
\end{bmatrix}$.\\
We then estimate the coefficients of the matrices $A_1$ and $A_2$ using our proposed method and the classical LS and Y-W method. For the FLOC-based estimators, we choose the parameter $B$ close to $\alpha -1$ as observed from Tables \ref{12}, \ref{11}, \ref{13} and \ref{15}, more specifically, $B=\alpha-1.05$. From Tables \ref{b}, \ref{c} and \ref{d}, we observe that the RMSE of the estimates obtained via the classical LS and Y-W method is fairly larger than our proposed FLOC method for different values of $n$ and $\alpha$. Summarising, our proposed method works better than classical LS and Y-W method.

\begin{table}[tbph]
	\centering
	\begin{tabular}{| *{7}{c |}}
		\hline
		$\textbf{B}$ &\textbf{0.00} & \textbf{0.11}& \textbf{0.22}&\textbf{0.33}&\textbf{0.44}&\textbf{0.55} \\
		\hline
		$a_1=0.1$& & & & & &\\
		\textbf{Mean}&0.1068 &0.0973 &0.1037 &0.0983 &0.1007 &0.0984\\
		\textbf{RMSE}&0.1634 &0.1244 &0.1156 &0.0923 &0.0850 &0.0856\\
		\hline
		$a_2=0.2$& & & & & &\\
		\textbf{Mean}&0.02057 &0.1735 &0.1983 &0.1862 &0.2016 &0.2017\\
		\textbf{RMSE}&0.3383 &0.3950 &0.2358 &0.2846 &0.1297 &0.1749\\
		\hline
		$a_3=0.3$& & & & & &\\
		\textbf{Mean}&0.3129 &0.3046 &0.3067 &0.3047 &0.3056 &0.2925\\
		\textbf{RMSE}&0.1980 &0.1887 &0.1768 &0.1754 &0.1320 &0.2206\\
		\hline
		$a_4=0.1$& & & & & &\\
		\textbf{Mean}&0.1288 &0.1141 &0.1173 &0.1198 &0.1216 &0.982\\
		\textbf{RMSE}&0.1654 &0.1343 &0.1144 &0.1642 &0.0912 &0.4188\\
		\hline
		$a_5=0.2$& & & & & &\\
		\textbf{Mean}&0.1750 &0.1841 &0.1834 &0.1912 &0.1945 &0.1937\\
		\textbf{RMSE}&0.1529 &0.1258 &0.1141 &0.0994 &0.0866 &0.0828\\
		\hline
		$a_6=0.05$& & & & & &\\
		\textbf{Mean}&0.0356 &0.0558 &0.0451 &0.0443 &0.0408 &0.0442\\
		\textbf{RMSE}&0.2655 &0.1969 &0.1767 &0.1966 &0.1020 &0.1524\\
		\hline
		$a_7=0.2$& & & & & &\\
		\textbf{Mean}&0.1947 &0.2007 &0.1959 &0.2062 &0.1913 &0.1998\\
		\textbf{RMSE}&0.1848 &0.1687 &0.1503 &0.1373 &0.1156 &0.1292\\
		\hline
		$a_8=0.1$& & & & & &\\
		\textbf{Mean}&0.0947 &0.1010 &0.0866 &0.1053 &0.0901 &0.1062\\
		\textbf{RMSE}&0.2515 &0.2027 &0.1393 &0.1292 &0.0890 &0.1679\\
		\hline
		
	\end{tabular}
	\caption{Mean and RMSE of 500 estimates of coefficients of $A_1$ and $A_2$ for $\alpha=1.6$ and $n=200$}
	\label{12}
\end{table}

\begin{table}[tbph]
	\centering
	\begin{tabular}{| *{7}{c |}}
		\hline
		$\textbf{B}$ &\textbf{0.00} & \textbf{0.11}& \textbf{0.22}&\textbf{0.33}&\textbf{0.44}&\textbf{0.55} \\
		\hline
		$a_1=0.1$& & & & & &\\
		\textbf{Mean}&0.0893 &0.0983 &0.0959 &0.1006 &0.1036 &0.0989\\
		\textbf{RMSE}&0.1479 &0.1000 &0.0772 &0.0578 &0.0500 &0.0393\\
		\hline
		$a_2=0.2$& & & & & &\\
		\textbf{Mean}&0.2410 &0.1896 &0.2022 &0.1951 &0.2034 &0.2021\\
		\textbf{RMSE}&0.9466 &0.4107 &0.1085 &0.1200 &0.0882 &0.0892\\
		\hline
		$a_3=0.3$& & & & & &\\
		\textbf{Mean}&0.3457 &0.3097 &0.3013 &0.2970 &0.3008 &0.3020\\
		\textbf{RMSE}&0.6879 &0.1852 &0.1230 &0.0791 &0.1711 &0.0768\\
		\hline
		$a_4=0.1$& & & & & &\\
		\textbf{Mean}&0.1374 &0.1153 &0.1199 &0.1241 &0.1203 &0.1270\\
		\textbf{RMSE}&0.5535 &0.0876 &0.0670 &0.0719 &0.0880 &0.0708\\
		\hline
		$a_5=0.2$& & & & & &\\
		\textbf{Mean}&0.1834 &0.1950 &0.1958 &0.1981 &0.1964 &0.1981\\
		\textbf{RMSE}&0.2322 &0.1022 &0.0743 &0.0596 &0.0587 &0.0429\\
		\hline
		$a_6=0.05$& & & & & &\\
		\textbf{Mean}&0.0164 &0.0480 &0.0416 &0.0424 &0.0404 &0.0403\\
		\textbf{RMSE}&0.6018 &0.2008 &0.0808 &0.0785 &0.0662 &0.0669\\
		\hline
		$a_7=0.2$& & & & & &\\
		\textbf{Mean}&0.1905 &0.1948 &0.2033 &0.1970 &0.2067 &0.1974\\
		\textbf{RMSE}&0.3064 &0.1265 &0.1024 &0.0720 &0.0850 &0.0575\\
		\hline
		$a_8=0.1$& & & & & &\\
		\textbf{Mean}&0.0660 &0.0996 &0.0972 &0.1003 &0.0977 &0.0950\\
		\textbf{RMSE}&0.5399 &0.1922 &0.0738 &0.0688 &0.0600 &0.0507\\
		\hline
	\end{tabular}
	\caption{Mean and RMSE of 500 estimates of coefficients of $A_1$ and $A_2$ for $\alpha=1.6$ and $n=700$}
	\label{11}
\end{table}

\begin{table}[tbph]
	\centering
	\begin{tabular}{| *{7}{c |}}
		\hline
		$\textbf{B}$ &\textbf{0.12} & \textbf{0.24}& \textbf{0.36}&\textbf{0.48}&\textbf{0.60}&\textbf{0.72} \\
		\hline
		$a_1=0.1$& & & & & &\\
		\textbf{Mean}&0.1023 &0.0965 &0.1019 &0.0978 &0.0988 &0.0987\\
		\textbf{RMSE}&0.1066 &0.0902 &0.0853 &0.0755 &0.0701 &0.0760\\
		\hline
		$a_2=0.2$& & & & & &\\
		\textbf{Mean}&0.2026 &0.1873 &0.1993 &0.1930 &0.2007 &0.1996\\
		\textbf{RMSE}&0.1461 &0.1499 &0.1215 &0.1464 &0.0909 &0.1037\\
		\hline
		$a_3=0.3$& & & & & &\\
		\textbf{Mean}&0.3040 &0.3038 &0.2968 &0.3017 &0.3026 &0.2918\\
		\textbf{RMSE}&0.1281 &0.1127 &0.1153 &0.1089 &0.0959 &0.1151\\
		\hline
		$a_4=0.1$& & & & & &\\
		\textbf{Mean}&0.1348 &0.1287 &0.1332 &0.1352 &0.1330 &0.1285\\
		\textbf{RMSE}&0.1109 &0.0996 &0.0924 &0.1072 &0.0805 &0.1611\\
		\hline
		$a_5=0.2$& & & & & &\\
		\textbf{Mean}&0.1855 &0.1886 &0.1892 &0.1922 &0.1965 &0.1940\\
		\textbf{RMSE}&0.1016 &0.0893 &0.0848 &0.0794 &0.0717 &0.0675\\
		\hline
		$a_6=0.05$& & & & & &\\
		\textbf{Mean}&0.0371 &0.0473 &0.0374 &0.0380 &0.0388 &0.0407\\
		\textbf{RMSE}&0.01315 &0.1044 &0.1102 &0.1152 &0.0804 &0.0951\\
		\hline
		$a_7=0.2$& & & & & &\\
		\textbf{Mean}&0.1951 &0.2009 &0.1952 &0.2042 &0.1930 &0.2002\\
		\textbf{RMSE}&0.1223 &0.1182 &0.1069 &0.1035 &0.0917 &0.0892\\
		\hline
		$a_8=0.1$& & & & & &\\
		\textbf{Mean}&0.0924 &0.0937 &0.0895 &0.0993 &0.0875 &0.0985\\
		\textbf{RMSE}&0.1221 &0.1090 &0.0927 &0.0849 &0.0721 &0.0836\\
		\hline
	\end{tabular}
		\caption{Mean and RMSE of 500 estimates of coefficients of $A_1$ and $A_2$ for $\alpha=1.75$ and $n=200$}
	\label{13}
\end{table}

\begin{table}[tbph]
	\centering
	\begin{tabular}{| *{7}{c |}}
		\hline
		$\textbf{B}$ &\textbf{0.12} & \textbf{0.24}& \textbf{0.36}&\textbf{0.48}&\textbf{0.60}&\textbf{0.72} \\
		\hline
		$a_1=0.1$& & & & & &\\
		\textbf{Mean}&0.0956 &0.0979 &0.0966 &0.0987 &0.1023 &0.0989\\
		\textbf{RMSE}&0.0618 &0.0578 &0.0527 &0.0438 &0.0411 &0.0342\\
		\hline
		$a_2=0.2$& & & & & &\\
		\textbf{Mean}&0.2031 &0.1948 &0.1996 &0.1977 &0.2014 &0.2012\\
		\textbf{RMSE}&0.2535 &0.1811 &0.0661 &0.0587 &0.0547 &0.0543\\
		\hline
		$a_3=0.3$& & & & & &\\
		\textbf{Mean}&0.3080 &0.3032 &0.2982 &0.2979 &0.3011 &0.3026\\
		\textbf{RMSE}&0.1766 &0.0810 &0.0664 &0.0529 &0.0919 &0.0512\\
		\hline
		$a_4=0.1$& & & & & &\\
		\textbf{Mean}&0.1319 &0.1308 &0.1336 &0.1382 &0.1344 &0.1380\\
		\textbf{RMSE}&0.0729 &0.0644 &0.0575 &0.0615 &0.0707 &0.0593\\
		\hline
		$a_5=0.2$& & & & & &\\
		\textbf{Mean}&0.1951 &0.1973 &0.1990 &0.1984 &0.1975 &0.1974\\
		\textbf{RMSE}&0.0828 &0.0585 &0.0486 &0.0442 &0.0423 &0.0362\\
		\hline
		$a_6=0.05$& & & & & &\\
		\textbf{Mean}&0.0405 &0.0424 &0.0404 &0.0388 &0.0387 &0.0382\\
		\textbf{RMSE}&0.1281 &0.1090 &0.0557 &0.0521 &0.0502 &0.0493\\
		\hline
		$a_7=0.2$& & & & & &\\
		\textbf{Mean}&0.2045 &0.1979 &0.2027 &0.1970 &0.2040 &0.1972\\
		\textbf{RMSE}&0.0877 &0.0679 &0.0614 &0.0500 &0.0565 &0.0449\\
		\hline
		$a_8=0.1$& & & & & &\\
		\textbf{Mean}&0.0898 &0.0945 &0.0951 &0.0978 &0.0956 &0.0923\\
		\textbf{RMSE}&0.1254 &0.0843 &0.0488 &0.0444 &0.0432 &0.0396\\
		\hline
	\end{tabular}
	\caption{Mean and RMSE of 500 estimates of coefficients of $A_1$ and $A_2$ for $\alpha=1.75$ and $n=700$}
	\label{15}
\end{table}

\begin{table}
	\centering
	\begin{tabular}{| *{8}{c |}}
		\hline
		\textbf{True Values} & \multicolumn{2}{| c }{\textbf{FLOC}}& \multicolumn{2}{| c  }{\textbf{LS}} &\multicolumn{2}{| c | }{\textbf{Y-W}}\\
		\hline
		$\boldsymbol{(n=100,\alpha=2, B=0.95)}$&  \textbf{Mean} & \textbf{RMSE} & \textbf{Mean}& \textbf{RMSE} & \textbf{Mean}&\textbf{RMSE} \\
		\hline
		$a_1=0.1$&0.088 &0.092 &0.072 &0.099 &0.079 &0.095\\
		\hline
		$a_2=0.2$&0.192 &0.104 &0.191 &0.119 &0.198 &0.100\\
		\hline
		$a_3=0.3$&0.297 &0.091 &0.290 &0.087 &0.297 &0.095\\
		\hline
		$a_4=0.1$&0.089 &0.084 &0.077 &0.011 &0.091 &0.096\\
		\hline
		$a_5=0.3$&0.277 &0.091 &0.251 &0.102 &0.251 &0.102\\
		\hline
		$a_6=0.4$&0.397 &0.097 &0.386 &0.090 &0.378 &0.094\\
		\hline
		$a_7=0.2$&0.204 &0.096 &0.193 &0.093 &0.192 &0.098\\
		\hline
		$a_8=0.1$&0.092 &0.081 &0.078 &0.103 &0.063 &0.102\\
		\hline
		& & & & & &\\
		\hline
		$\boldsymbol{(n=800,\alpha=2, B=0.95)}$&  \textbf{Mean} & \textbf{RMSE} & \textbf{Mean}& \textbf{RMSE}& \textbf{Mean}&\textbf{RMSE}  \\
		\hline
		$a_1=0.1$&0.096 &0.032 &0.096 &0.033 &0.095 &0.033\\
		\hline
		$a_2=0.2$&0.198 &0.036 &0.203 &0.033 &0.200 &0.035\\
		\hline
		$a_3=0.3$&0.300 &0.032 &0.296 &0.029 &0.300 &0.032\\
		\hline
		$a_4=0.1$&0.098 &0.028 &0.094 &0.030 &0.096 &0.032\\
		\hline
		$a_5=0.3$&0.298 &0.032 &0.299 &0.032 &0.296 &0.033\\
		\hline
		$a_6=0.4$&0.400 &0.032 &0.401 &0.025 &0.399 &0.032\\
		\hline
		$a_7=0.2$&0.201 &0.033 &0.202 &0.029 &0.200 &0.034\\
		\hline
		$a_8=0.1$&0.098 &0.027 &0.095 &0.034 &0.095 &0.033\\
		\hline
	\end{tabular}
		\caption{Comparison of mean and RMSE of 500 estimates of coefficients of $A_1$ and $A_2$ using FLOC, LS and Y-W method}
	\label{b}
\end{table}

\begin{table}
	\centering
	\begin{tabular}{| *{8}{c |}}
		\hline
		\textbf{True Values} & \multicolumn{2}{| c }{\textbf{FLOC}}& \multicolumn{2}{| c | }{\textbf{LS}} &\multicolumn{2}{| c | }{\textbf{Y-W}} \\
		\hline
		$\boldsymbol{(n=200,\alpha=1.85, B=0.8)}$&  \textbf{Mean} & \textbf{RMSE} & \textbf{Mean}& \textbf{RMSE} & \textbf{Mean}&\textbf{RMSE} \\
		\hline
		$a_1=0.1$&0.098 &0.068 &0.092 &0.066 &0.096 &0.066\\
		\hline
		$a_2=0.2$&0.195 &0.081 &0.192 &0.076 &0.195 &0.078\\
		\hline
		$a_3=0.3$&0.300 &0.073 &0.0296 &0.070 &0.299 &0.076\\
		\hline
		$a_4=0.1$&0.087 &0.069 &0.098 &0.060 &0.100 &0.060\\
		\hline
		$a_5=0.3$&0.286 &0.069 &0.279 &0.070 &0.272 &0.074\\
		\hline
		$a_6=0.4$&0.403 &0.073 &0.396 &0.067 &0.391 &0.070\\
		\hline
		$a_7=0.2$&0.196 &0.073 &0.194 &0.069 &0.190 &0.071\\
		\hline
		$a_8=0.1$&0.098 &0.059 &0.085 &0.066 &0.081 &0.068\\
		\hline
		& & & & & &\\
		\hline
		$\boldsymbol{(n=700,\alpha=1.85, B=0.8)}$&  \textbf{Mean} & \textbf{RMSE} & \textbf{Mean}& \textbf{RMSE}& \textbf{Mean}&\textbf{RMSE}  \\
		\hline
		$a_1=0.1$&0.098 &0.035 &0.097 &0.034 &0.098 &0.034\\
		\hline
		$a_2=0.2$&0.198 &0.067 &0.197 &0.044 &0.197 &0.045\\
		\hline
		$a_3=0.3$&0.299 &0.039 &0.298 &0.038 &0.302 &0.060\\
		\hline
		$a_4=0.1$&0.085 &0.059 &0.100 &0.031 &0.100 &0.032\\
		\hline
		$a_5=0.3$&0.296 &0.035 &0.294 &0.035 &0.292 &0.036\\
		\hline
		$a_6=0.4$&0.404 &0.052 &0.399 &0.037 &0.397 &0.032\\
		\hline
		$a_7=0.2$&0.202 &0.043 &0.201 &0.043 &0.200 &0.047\\
		\hline
		$a_8=0.1$&0.103 &0.042 &0.094 &0.035 &0.093 &0.036\\
		\hline
	\end{tabular}
	\caption{Comparison of mean and RMSE of 500 estimates of coefficients of $A_1$ and $A_2$ using FLOC, LS and Y-W method}
	\label{c}
\end{table}
\begin{table}
	\centering
	\begin{tabular}{| *{8}{c |}}
		\hline
		\textbf{True Values} & \multicolumn{2}{| c }{\textbf{FLOC}}& \multicolumn{2}{| c | }{\textbf{LS}}  &\multicolumn{2}{| c | }{\textbf{Y-W}}\\
		\hline
		$\boldsymbol{(n=300,\alpha=1.65, B=0.6)}$&  \textbf{Mean} & \textbf{RMSE} & \textbf{Mean}& \textbf{RMSE} & \textbf{Mean}&\textbf{RMSE} \\
		\hline
		$a_1=0.1$&0.098 &0.059 &0.095 &0.081 &0.095 &0.051\\
		\hline
		$a_2=0.2$&0.200 &0.109 &0.199 &0.073 &0.199 &0.071\\
		\hline
		$a_3=0.3$&0.298 &0.073 &0.295 &0.078 &0.295 &0.079\\
		\hline
		$a_4=0.1$&0.056 &0.118 &0.093 &0.053 &0.094 &0.038\\
		\hline
		$a_5=0.3$&0.293 &0.062 &0.292 &0.054 &0.287 &0.059\\
		\hline
		$a_6=0.4$&0.410 &0.093 &0.395 &0.073 &0.389 &0.093\\
		\hline
		$a_7=0.2$&0.195 &0.122 &0.190 &0.181 &0.183 &0.0184\\
		\hline
		$a_8=0.1$&0.116 &0.077 &0.092 &0.052 &0.089 &0.055\\
		\hline
		& & & & & &\\
		\hline
		$\boldsymbol{(n=600,\alpha=1.65, B=0.6)}$&  \textbf{Mean} & \textbf{RMSE} & \textbf{Mean}& \textbf{RMSE}& \textbf{Mean}&\textbf{RMSE}  \\
		\hline
		$a_1=0.1$&0.101 &0.041 &0.097 &0.034 &0.099 &0.035\\
		\hline
		$a_2=0.2$&0.201 &0.083 &0.200 &0.094 &0.198 &0.058\\
		\hline
		$a_3=0.3$&0.301 &0.047 &0.301 &0.041 &0.300 &0.045\\
		\hline
		$a_4=0.1$&0.053 &0.107 &0.100 &0.033 &0.100 &0.034\\
		\hline
		$a_5=0.3$&0.297 &0.040 &0.296 &0.032 &0.294 &0.037\\
		\hline
		$a_6=0.4$&0.412 &0.064 &0.397 &0.065 &0.396 &0.043\\
		\hline
		$a_7=0.2$&0.197 &0.048 &0.195 &0.040 &0.196 &0.037\\
		\hline
		$a_8=0.1$&0.116 &0.066 &0.095 &0.043 &0.095 &0.036\\
		\hline
	\end{tabular}
	\caption{Comparison of mean and RMSE of 500 estimates of coefficients of $A_1$ and $A_2$ using FLOC, LS and Y-W method}
	\label{d}
\end{table}
\newpage
\section{Application to Financial Data}\label{app}
In this section, we give an empirical analysis of the real financial bivariate data using the vector autoregressive model with symmetric stable noise. We consider the monthly simple returns of the stocks of International Business Machines (IBM) and the SP Composite index (SP) from January 1961 to December 2011 with 612 observations. The datasets are available in the ``MTS" package in R.

\noindent
To determine whether the considered time series datasets are stationary, we implement the Augmented Dickey-Fuller Test (ADF) ``adf.test" available in the ``tseries" package in R. The $p$-values obtained are 0.01 which confirms the stationarity of the datasets. Next, we fit our proposed FLOC based bidimensional autoregressive model with $p=2$ to the data. Since $A=1$ is fixed, we need to estimate $\alpha$ from the bivariate data to find the value of $B$. We use the hybrid method \cite{hybrid} to obtain the estimate of $\alpha$. Thus, $\hat{\alpha}=1.86$ and $B=0.8$, value close to $1-\hat{\alpha}$ as observed in the simulation study. The estimates of the coefficient matrices are as follows:
 $\widehat{A}_1=\begin{bmatrix}
0.003 & 0.069\\
0.014 & 0.023
\end{bmatrix}$ and $\widehat{A}_2=\begin{bmatrix}
-0.040 & 0.031\\
0.021 & 0.020
\end{bmatrix}$.\\
Next, in order to check if the fitted model is appropriate we analyse the residuals of the time series $\lbrace X_{1t}\rbrace$ and $\lbrace X_{2t}\rbrace$. In our model, we assume that the noise series $\lbrace Z_{1t}\rbrace$ or $\lbrace Z_{2t}\rbrace$ is a sample of independent and stable distributed random variables. Thus, we fit stable distribition using the hybrid method \cite{hybrid} to the residual time series $\lbrace Z_{1t}\rbrace$ and $\lbrace Z_{2t}\rbrace$. The estimates obtained are ($\hat{\alpha}, \hat{\beta}, \hat{\sigma}, \hat{\delta})=(1.900, -0.708, 0.027, 0.008)$ and $1.851, 0.148, 0.035, 0.010$), respectively for  $\lbrace Z_{1t}\rbrace$ and $\lbrace Z_{2t}\rbrace$.

\begin{figure*}
     \caption{\textbf{Residuals and Auto-FLOC Plot of $Z_{1t}$}} \label{fig1}
	\begin{multicols}{2}
		\includegraphics[width=\linewidth]{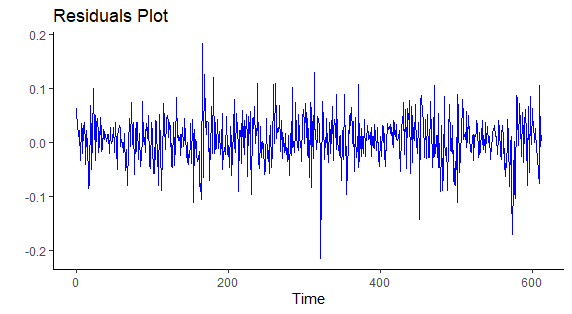}\par 
		\includegraphics[width=\linewidth]{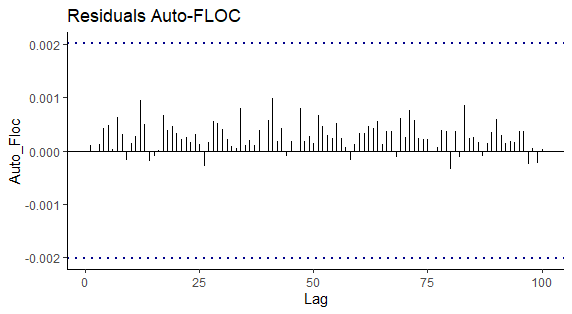}\par 
	\end{multicols}
\end{figure*}

\begin{figure*}
	\caption{\textbf{Residuals and Auto-FLOC Plot of $Z_{2t}$}} \label{fig2}
	\begin{multicols}{2}
		\includegraphics[width=\linewidth]{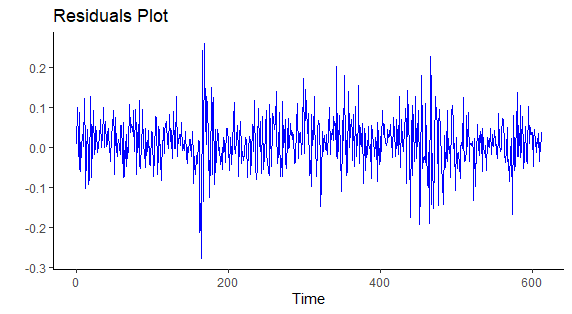}\par 
		\includegraphics[width=\linewidth]{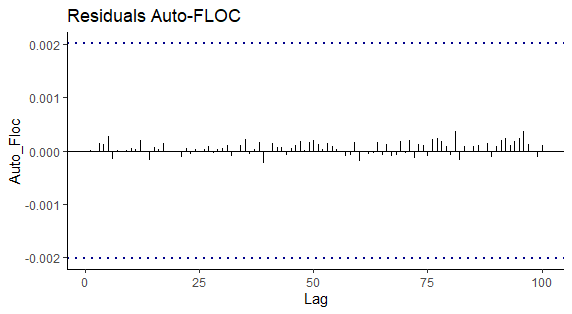}\par 
	\end{multicols}
\end{figure*}

\begin{figure*}
	\caption{\textbf{QQPlots of $Z_{1t}$ and $Z_{2t}$}} \label{fig3}
	\begin{multicols}{2}
		\includegraphics[width=\linewidth]{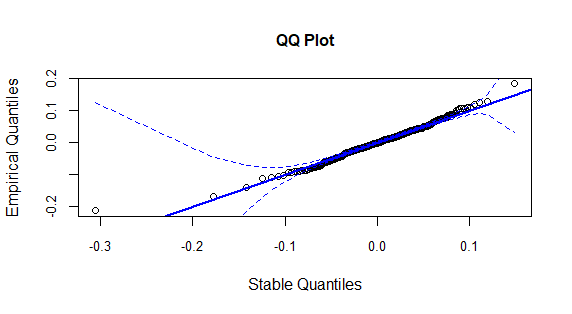}\par 
		\includegraphics[width=\linewidth]{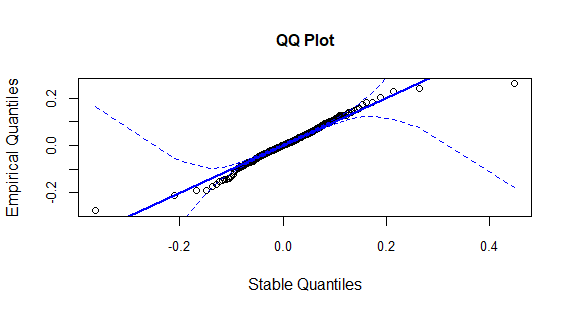}\par 
	\end{multicols}
\end{figure*}

\noindent
In order to check if the residuals can be considered as an independent sample, we make use of the empirical auto-FLOC function defined in \cite{p} instead of the classical autocovariance (autocorrelation) function for our model. From the residuals auto-FLOC plot in Figure \ref{fig1} and \ref{fig2}, one can observe that the dependence in the residual series is almost unidentifiable. Thus, we assume that  $\lbrace Z_{1t}\rbrace$ and $\lbrace Z_{2t}\rbrace$ are independent for all values of $t$. Note that for empirical auto-FLOC function for  $\lbrace Z_{1t}\rbrace$, $A=1$ and $B=0.85$ while for  $\lbrace Z_{2t}\rbrace$, $A=1$ and $B=0.8$. Additionally, the QQ plots as shown in Figure \ref{fig3}, represent the quantiles for fitted stable distribution with the estimated parameters from the residuals and the empirical quantiles for residuals. We observe the distribution of both the residuals is almost stable. Finally, we perform the Kolmogorov-Smirnov (KS) test as discussed in \cite{yule} with the hypothesis that $\lbrace Z_{1t}\rbrace$ and $\lbrace Z_{2t}\rbrace$ have stable distribution. Since the obtained $p$-values are 0.60 and 0.7, respectively for $\lbrace Z_{1t}\rbrace$ and $\lbrace Z_{2t}\rbrace$, calculated on the basis of 100 Monte Carlo repetitions, we fail to reject the hypothesis at the significance level 0.05.

\newpage
\section{Conclusion}\label{con}
To conclude, we make the following observations in relation to our proposed methods.
\begin{itemize}
\item In this article, a new estimation method for multidimensional VAR($p$), $p \geq 1$ with symmetric stable distribution ($\alpha \in [1.5, 2])$ is introduced. 
\item The proposed method is an extension of the classical Y-W method which is based on the covariance function of the underlying process.
\item The simulation study reflects the effectiveness of the proposed method in comparison to the classical LS and Y-W method.  The application of the FLOC measure is justified from the theoretical point of view in the considered case (the theoretical covariance does not exist) however
by simulation study we have proved it is reasonable to
use the new technique taking under account the practical
aspects. 
\item Finally, we fit our proposed model to the bivariate financial data.
\end{itemize}
\newpage
\tiny

\end{document}